\documentstyle[epsfig]{aipproc}

\begin{document}
\title{Evidence of $\phi\to\pi^0\pi^0\gamma$ and
$\phi\to\pi^0\eta\gamma$ decays in SND experiment at VEPP-2M}
\author{
M.N.Achasov,
A.V.Berdyugin, 
A.V.Bozhenok,
A.D.Bukin,
D.A.Bukin,
S.V.Burdin,
T.V.Dimova,
V.P.Druzhinin,
M.S.Dubrovin,
I.A.Gaponenko,
V.B.Golubev,
V.N.Ivanchenko,
I.A.Koop,
A.A.Korol,
S.V.Koshuba,
E.V.Pakhtusova,
A.A.Salnikov,
S.I.Serednyakov,
V.V.Shary,
Yu.M.Shatunov,
V.A.Sidorov, 
Z.K.Silagadze,
A.N.Skrinsky,
Yu.V.Usov
}
\address{Budker Institute of Nuclear Physics, Novosibirsk, 630090, Russia}

\maketitle

\begin{abstract}
   Preliminary results on the study of $e^+e^-
\to\phi(1020)\to\pi^0\pi^0\gamma$,
$\eta\pi^0\gamma$ processes from
SND experiment at VEPP-2M collider in Novosibirsk
are presented. Branching ratios of rare radiative $\phi\to\pi^0\pi^0\gamma$
and $\phi\to\pi^0\eta\gamma$ decays are measured:

$B(\phi \to \pi^o\pi^o\gamma ) = (1.1 \pm 0.2) \cdot 10^{-4}~ 
(M_{\pi\pi} < 800 MeV)$,
 
$B(\phi \to\eta \pi^o\gamma ) = (1.3 \pm 0.5) \cdot 10^{-4}$.

\end{abstract}

First upper limits of
$\phi(1020)\to\pi^o\pi^o\gamma,\eta\pi^o\gamma$ radiative decays were 
established in ND experiment [1] at  VEPP-2M $e^+ e^-$ collider.
Later it was shown by N.N.Achasov [2] that these
decays can provide important information about quark
structure of lightest scalar mesons
$a_0(980)$ and $f_0(980)$. Further theoretical investigations
[3-9] confirmed this idea.

In this work we present preliminary results 
of  SND experiment [10,11].
Main background for the decays under study is due to
$\phi\to\eta\gamma\to3\pi^o\gamma$ and
$\phi \to K_SK_L \to \pi^0\pi^0K_L$ decays. To suppress 
it, events with 5 photons
were selected, satisfying energy-momentum conservation.
In addition a $\chi^2_{\gamma}$ parameter, quantitatively
describing
quality of photons [12], was used (Fig.1).
The spectrum of invariant masses of photon pairs
in selected events (Fig.2a) shows that
most of them contain two $\pi^0$.
The  $\pi^0\gamma$ invariant 
mass distribution (Fig.2b) was used to separate
the decay under study from the remaining
$e^+e^-\to\omega\pi^0\to\pi^0\pi^0\gamma$
background (Table~1).  
The polar angle distributions of $\pi^0$-s in
$\pi^0\pi^0$ pair rest frame are shown in Figs.2c,d.
Flat distribution of $f_0\gamma$-type events (Fig.2c) is
consistent with $S$-wave production mechanism, while
distribution of $\omega\pi^0$-type events (Fig.2d) clearly
contradicts it.

\begin{table}
\caption{Number of the selected $\pi^o\pi^o\gamma$ events with
$m_{\pi^o \pi^o} > 800~MeV$ and estimated background.}
\label{IVN:T1}
\begin{tabular}{ll}
$\phi \to \pi^o \pi^o \gamma $, experiment  & 45 \\
$\phi \to \eta \gamma $, simulation  & 5 \\
$\phi \to K_SK_L $, simulation  & $< 6$ \\
$\phi \to \rho \pi^o, \omega \pi^o \to \pi^o \pi^o \gamma $
 , simulation & 1.4 \\
$B(\phi \to \pi^o \pi^o \gamma )$  & $(1.1 \pm 0.2)\cdot 10^{-4}$ \\
\end{tabular}
\end{table}

The $\pi^o \pi^o $ invariant
mass spectrum, corrected for detection efficiency 
dependence on $m_{\pi\pi}$, (Fig.3)
shows visible peak close to $f_0$ mass. Its width 
$(\sim60~MeV)$ and position $(\sim950~MeV)$
do not contradict previous measurements [13]. 
The shape of the invariant mass spectrum is consistent with
4-quark model predictions [2,9], which allows us to fit it  
using formulas of Ref.[2]. 
The results are the following:

$m_{f_0} = (950 \pm 8)~MeV$, $g^2_{f\pi\pi}/4\pi = (0.4 \pm 0.1)~GeV^{-2}$,

$B(\phi \to f_0(980) \gamma) = (4.7 \pm 1.0)\cdot 10^{-4}$.

Situation with $\phi\to\eta\pi^0\gamma$  decay mode is
less clear because there is no visible peak at $\eta$ mass in 
the spectrum in Fig.2a. But the number of selected $\eta\pi^0\gamma$
candidate events significantly exceeds
estimated background (Figs.4a,b). To check consistency of our analysis, 
3 sets
of selection criteria were used (Table 2). Observed background in the
kinematic region, where only small number of
$\eta\pi^0\gamma$ events were expected, is well
reproduced by Monte Carlo simulation (Fig.4d), indicating that our 
estimations of  background in the ``effect'' kinematic region are correct. 
Soft (Fig.4a) and strong (Fig.4b) cuts give practically the same mass 
spectra, final mass spectrum after background subtraction is shown 
in Fig.4c.

\begin{table}
\caption{Number of the selected $\eta\pi^0\gamma$ events
 and estimated background.}
\label{IVN:T2}
\begin{tabular}{llll}
$\chi^2_{\gamma}$              &  $<$0 (Strong cuts)   &  
$<$25 (Soft cuts)              &  25-50 (Background)     \\
\tableline
$e^+e^-\to\eta\pi\gamma$, experiment   & 34     &  283      &  96        \\
$e^+e^-\to\eta\gamma$, simulation      & 4.2    &  109      &  50        \\
$e^+e^- \to K_SK_L $, simulation       & $<$5.8 &   10      &  20        \\
$e^+e^- \to\omega\pi$, simulation      & 2.1    &   85      &   4        \\
\tableline
$B(\phi\to\eta\pi^0\gamma)$    & $(1.5\pm0.5)\cdot10^{-4}$ &
$(1.3\pm0.5)\cdot10^{-4}$      & $(0.6\pm0.5)\cdot10^{-4}$             \\
\end{tabular}
\end{table}

In conclusion we would like to emphasize that SND measurement of 
$B(\phi\to\eta\gamma)$ in  $7\gamma$ final state
 (Table 9 in Ref.[11]) is in a good agreement with the PDG value [13] and
upper limits, presented in the same table for several forbidden
neutral decay modes are established at levels significantly lower than
branching ratios measured in this work:    

$B(\phi \to \pi^o\pi^o\gamma ) = (1.1 \pm 0.2) \cdot 10^{-4}~ 
(M_{\pi\pi} < 800 MeV)$,

$B(\phi \to\eta \pi^o\gamma ) = (1.3 \pm 0.5) \cdot 10^{-4}$.

This analysis is practically model independent and its results are 
more in favor of 4-quark model.
Further theoretical studies are necessary to prove this conclusion as well as
additional experimental data are needed to confirm our 
observation of $\phi\to\pi^0\pi^0\gamma$ and $\phi\to\eta\pi^0\gamma$ decays.

{\bf Acknowledgment}.
The work is partially supported by RFFR.
\mbox{Grant No 96-02-19192.}

\begin{figure}[b!] 
\centerline{\epsfig{file=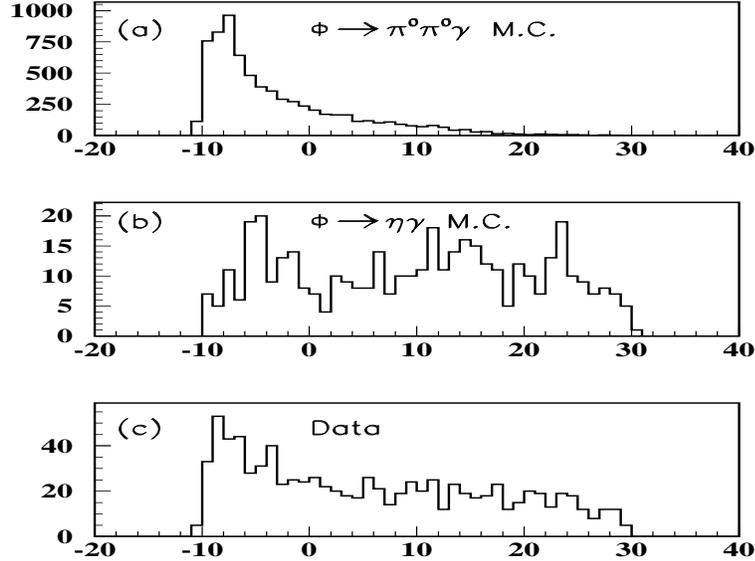,height=3.5in,width=4.5in}}
\vspace{5pt}
\caption{The photon likelihood function $\chi^2_{\gamma}$ for 
Monte Carlo (a,b) and for the data (c).}
\label{fig1}
\end{figure}

\begin{figure}[b!] 
\centerline{\epsfig{file=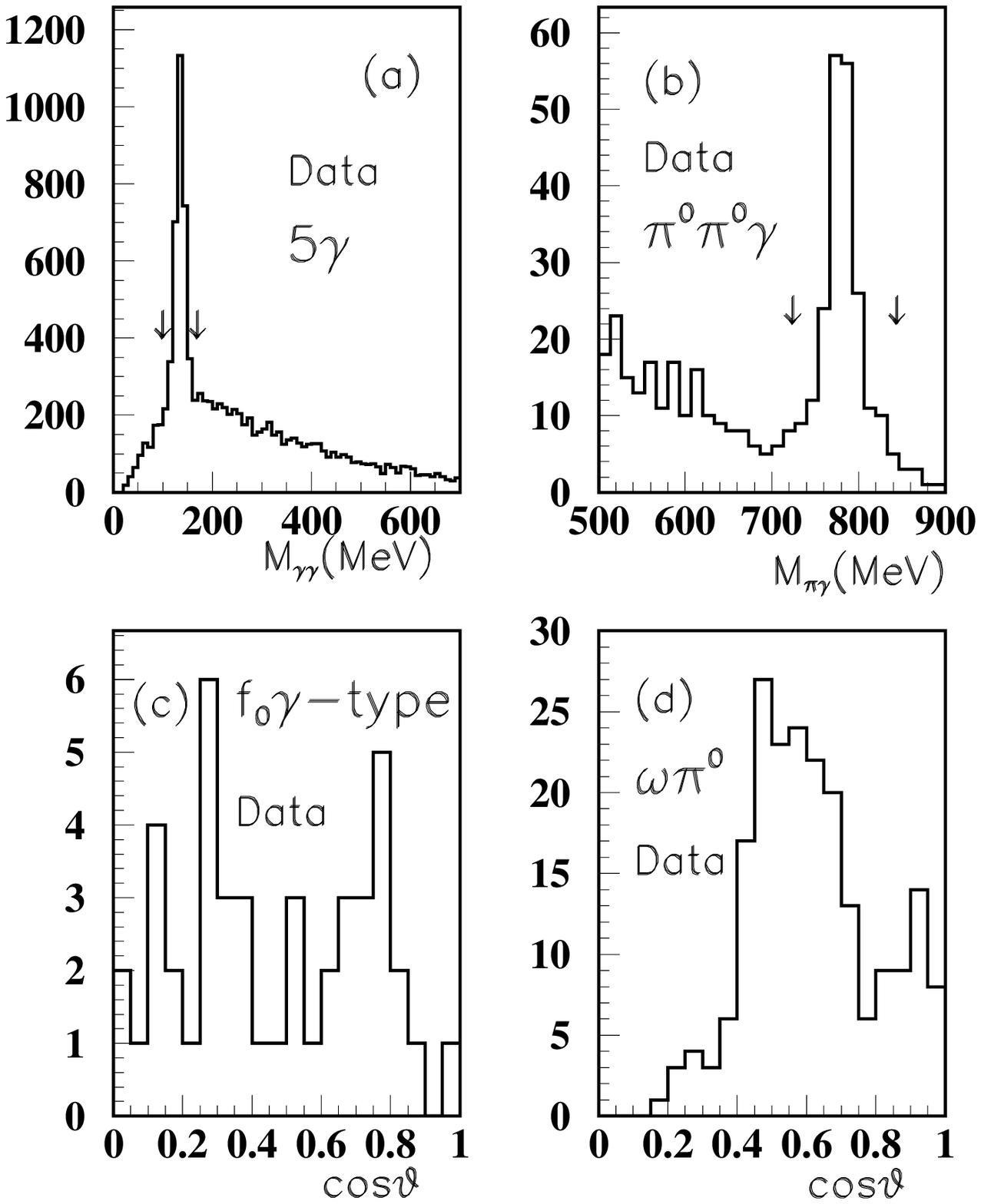,height=3.8in,width=5.5in}}
\vspace{1pt}
\caption{Selection of $\pi^0\pi^0\gamma$ events.}
\label{fig2}
\end{figure}

\begin{figure}[b!] 
\centerline{\epsfig{file=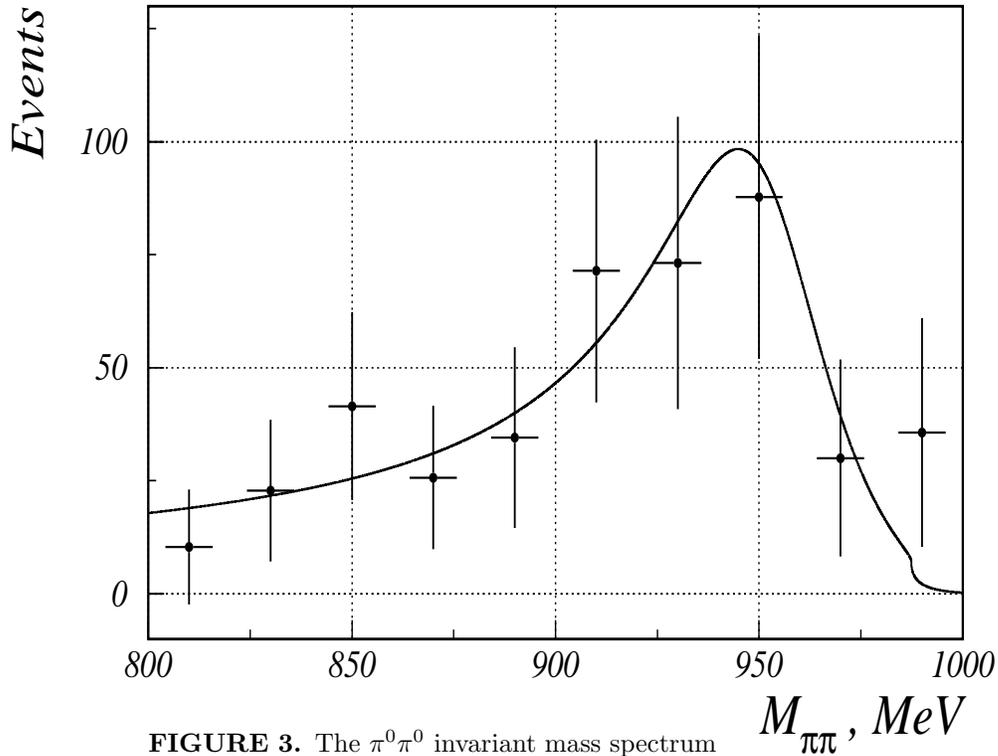,height=3.8in,width=5.5in}}
\vspace{1pt}
\caption{The $\pi^0\pi^0$ invariant mass spectrum \qquad \qquad \qquad}
\label{fig3}
\end{figure}

\begin{figure}[b!] 
\centerline{\epsfig{file=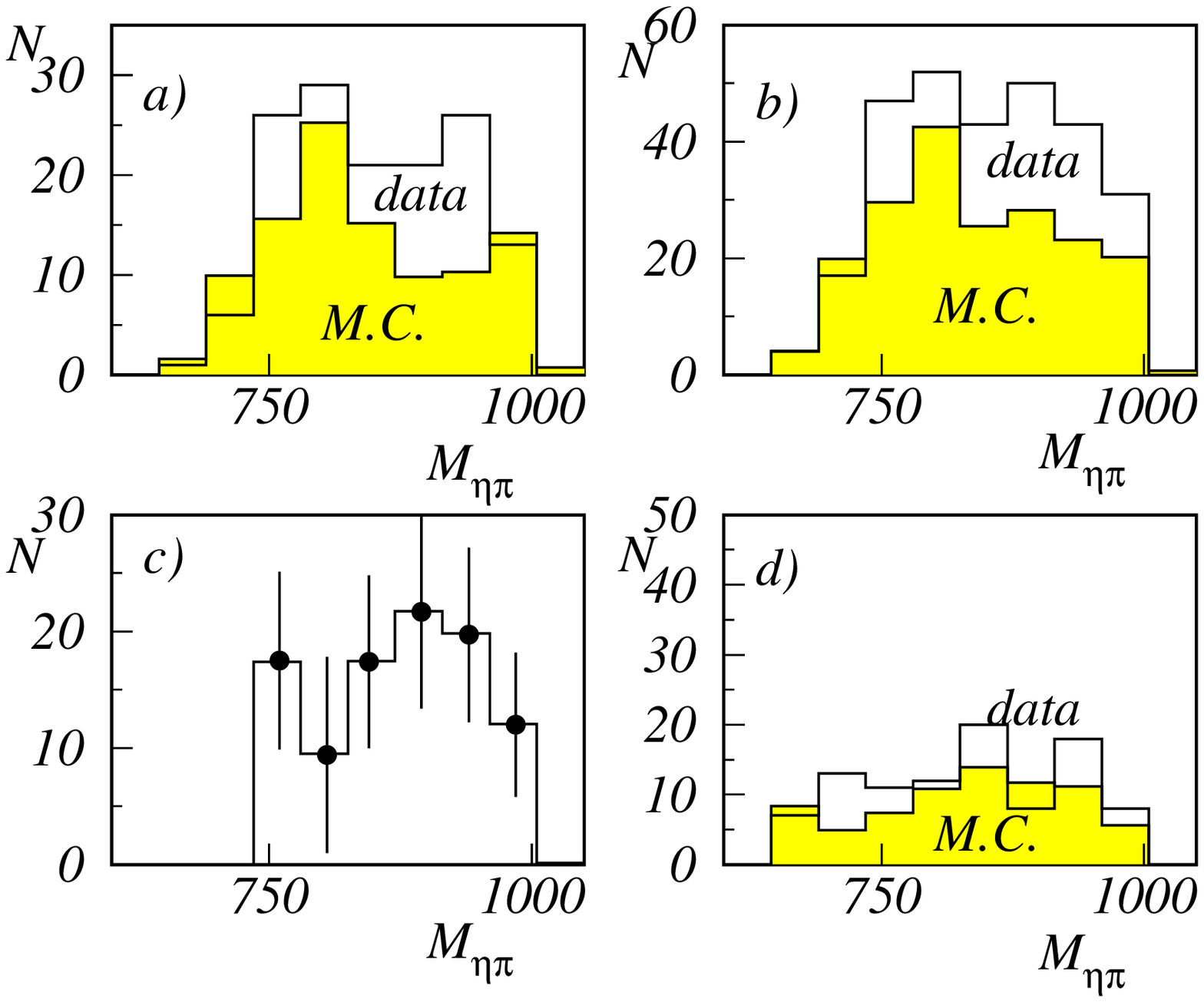,height=3.8in,width=5.5in}}
\vspace{1pt}
\caption{Selection of $\eta\pi^0\gamma$ events.}
\label{fig4}
\end{figure}

\end{document}